\documentclass[article,pdftex,shortnames]{jss}\usepackage[]{graphicx}\usepackage[]{color}
\makeatletter
\def\maxwidth{ %
  \ifdim\Gin@nat@width>\linewidth
    \linewidth
  \else
    \Gin@nat@width
  \fi
}
\makeatother

\definecolor{fgcolor}{rgb}{0.345, 0.345, 0.345}

\usepackage{framed}
\makeatletter
\newenvironment{kframe}{%
 \def\at@end@of@kframe{}%
 \ifinner\ifhmode%
  \def\at@end@of@kframe{\end{minipage}}%
  \begin{minipage}{\columnwidth}%
 \fi\fi%
 \def\FrameCommand##1{\hskip\@totalleftmargin \hskip-\fboxsep
 \colorbox{shadecolor}{##1}\hskip-\fboxsep
     \hskip-\linewidth \hskip-\@totalleftmargin \hskip\columnwidth}%
 \MakeFramed {\advance\hsize-\width
   \@totalleftmargin\z@ \linewidth\hsize
   \@setminipage}}%
 {\par\unskip\endMakeFramed%
 \at@end@of@kframe}
\makeatother

\definecolor{shadecolor}{rgb}{.97, .97, .97}
\definecolor{messagecolor}{rgb}{0, 0, 0}
\definecolor{warningcolor}{rgb}{1, 0, 1}
\definecolor{errorcolor}{rgb}{1, 0, 0}
\newenvironment{knitrout}{}{} % an empty environment to be redefined in TeX

\usepackage{alltt}

\usepackage[ruled,noline,linesnumbered]{algorithm2e}
\SetKwFor{For}{for}{do}{end}
\SetKwFor{While}{while}{do}{end}
\SetKwInput{KwIn}{input}
\SetKwInput{KwOut}{output}
\SetKwInput{KwCplx}{complexity}
\SetKwBlock{Begin}{procedure}{}
\DontPrintSemicolon

\usepackage{paralist}
\usepackage{amsmath}
\usepackage{amssymb}
\usepackage{color}
\usepackage{graphicx}
\usepackage[all]{hypcap}

\usepackage[sort&compress]{cleveref}

\crefname{figure}{Fig.}{Figs.}
\Crefname{figure}{Fig.}{Figs.}
\crefname{table}{Table}{Tables}
\Crefname{table}{Table}{Tables}
\crefname{equation}{Eq.}{Eqs.}
\Crefname{equation}{Eq.}{Eqs.}
\creflabelformat{equation}{#2#1#3}
\crefname{appendix}{Appendix}{Appendices}
\Crefname{appendix}{Appendix}{Appendices}
\crefname{algorithm}{Algorithm}{Algorithms}
\Crefname{algorithm}{Algorithm}{Algorithms}
\crefname{AlgoLine}{line}{lines}
\Crefname{AlgoLine}{Line}{Lines}

\newtheorem{theorem}{Theorem}

\newcommand\loglik{\ell}
\newcommand\loglikMC{\hat\ell}

\newcommand\class[1]{class `\textrm{#1}'}
\newcommand\prob[1]{\mathbb{P}\left[{#1}\right]}

\newcommand\given{{\,\vert\,}}

\newcommand\myto{{\;:\;}}
\newcommand\seq[2]{{#1}\!:\!{#2}}
\newcommand\mydot{{\,\cdot\,}}

\newcommand\dlta[1]{\Delta{#1}}
\newcommand\giventh{{\hspace{0.5mm};\hspace{0.5mm}}}
\newcommand\normal{{\mathrm{Normal}}}
\newcommand\argequals{{\,=\,}}

\newcommand\bigO[1]{\mathcal{O}\!\left({#1}\right)}

\author{Duc Anh Doan, Dao Nguyen, Xin Dang\\University of Mississippi\\}
\title{Simulation-based inference methods for partially observed Markov model via the \proglang{R} package \pkg{is2}}
\Plainauthor{Duc Anh Doan, Dao Nguyen, Xin Dang} %% comma-separated
\Plaintitle{Simulation-based inference methods} %% without formatting
\Shorttitle{Simulation-based inference methods}

\Abstract{
Partially observed Markov process (POMP) models are powerful tools for time series modeling and analysis. 
Inherited the flexible framework of R package \pkg{pomp}, the \pkg{is2} package extends some useful 
Monte Carlo statistical methodologies to improve on convergence rates. A variety of efficient statistical methods for POMP
models have been developed including fixed lag smoothing, second-order iterated smoothing, momentum iterated filtering,
average iterated filtering, accelerate iterated filtering and particle iterated filtering. 
In this paper, we show the utility of these methodologies based on two
toy problems. We also demonstrate the potential of some methods in a more complex model, 
employing a nonlinear epidemiological model with a discrete population, seasonality, and
extra-demographic stochasticity. We discuss the extension beyond POMP models and
the development of additional methods within the framework provided by
\pkg{is2}.
}
\Plainkeywords{Markov processes, state space model, stochastic dynamical system, maximum likelihood, fixed lag smoothing, R}
\Address{
  Duc Anh Doan\\
  Department of Mathematics\\
  University of Mississippi\\
  38677 Mississippi, USA\\
  E-mail: \email{ddoan@go.olemiss.edu}\\
  
  Dao Nguyen\\
  Department of Mathematics\\
  University of Mississippi\\
  38677 Mississippi, USA\\
  E-mail: \email{dxnguyen@olemiss.edu}\\
  URL: \url{https://math.olemiss.edu/dao-nguyen/}\\

  Xin Dang\\
  Department of Mathematics\\
  University of Mississippi\\
  38677 Mississippi, USA\\
  E-mail: \email{xdang@olemiss.edu}\\
  URL: \url{home.olemiss.edu/~xdang/}\\

}
\IfFileExists{upquote.sty}{\usepackage{upquote}}{}
\begin{document}

%% include your article here, just as usual
%% Note that you should use the \pkg{}, \proglang{} and \code{} commands.

\section {Introduction}
In data analysis, partially observed
Markov process (POMP) models (also known as state space models) have
been known of as powerful tools in modeling and analyzing time series in many disciplines
such as ecology, econometric, engineering and statistics. However, making inferences
on POMP models can be problematic because of the presence of incomplete
measurements, and possibly weakly identifiable parameters. Typical
methods for inference (e.g., maximum likelihood) with strong
assumptions of linear Gaussian models have often resulted in undesired
outcomes, especially when the assumptions are violated. Simulation-based inferences,
also called plug-and-play inferences, are characterized as the dynamic model
entering the statistical computations through a numerical solution of the sample
paths \citep{breto09}. As a result, there has been a lot of
interest in simulation-based inference for POMP models, largely
software development for this class of inference. The \pkg{pomp}
software package \citep{pomp} was developed to provide a general
representation of a POMP model where several algorithms
implemented within \pkg{pomp} are applicable to arbitrary POMP
models \citep{ionides06,toni09,andrieu10,wood10,ionides15, nguyen2017second}.
However, since \pkg{pomp} is designed years ago for conventional POMP
models, applicabilities of these inference methods for very large-scale
data set remains an open question. In particular, in the big data
regime when the convergence rate is critical, some recently efficient methods
have not yet been parts of \pkg{pomp}. Motivated by this need and the popularity
of \pkg{pomp}, we develop \pkg{is2} software package to provide additional efficient algorithms. 
Based on the recent developments of stochastic optimization literature, we
focus more on inference methods using advanced machine learning algorithms, which we consider as 
a complement of \pkg{pomp} rather than a competing software
package.

Specifically, we extend the core functionality of \pkg{pomp} to include particle
fixed lag smoothing which is more stable than filtering and reduces computational expense of
traditional smoothing. In addition, we provide several efficient inference methods using the full
information and full likelihood of the data. 
The first algorithm developed to carry out such inference
is the second order iterated smoothing algorithm of \citet{nguyen2017second}. 
\citet{doucet2013derivative} showed that sequential
Monte Carlo smoothing can give the first derivative as well as Hessian estimation
of the log likelihood that have better asymptotic convergence rates
than those used for iterated filtering in \pkg{pomp}. One can apply
these moments approximations of \citet{doucet2013derivative} in a Newton-Raphson
type optimization algorithm. We 
develop a modification of the theory of \citet{doucet2013derivative} giving
rise to a new algorithm, which empirically shows
enhanced performance over other available methods on standard benchmarks. The second implemented algorithm is called momentum iterated filtering where
we accumulate a velocity vector in the persistent increase directions of the log likelihood function across iterations.
This will help the algorithm achieve results of the same level of accuracy in fewer iterations. 
The third algorithm developed
for simulation-based inference is accelerate iterated filtering. Unlike
original iterated filtering, the accelerate method exploits an approximation
of the gradient of log likelihood in the accelerated scheme where
the convexity and unbiasedness restrictions are relaxed. The theoretical underlying is described
in more details in \citep{nguyen2018accelerate}. By avoiding the computational expenses
of Hessian, this method is very promising as it is a first order but
obtains the quadratic convergence rate of the second order approaches. The fourth
algorithm included in the package is average iterated filtering,
which has typically been motivated by two time scales optimization
process. In this algorithm, the slower scale will be sped up by the
faster scale, resulting in a very attractive approach with a simple
computation but a fast convergence rate. 
Lastly, we implement a particle
version of iterated filtering algorithm where gradient of the log
likelihood as the by-product of iterated filtering is used to improve
the proposal distribution in Langevin algorithm in the context of simulation-based
algorithm. This algorithm enables a routine full-information plug-and-play
Bayesian inference for POMP models, inheriting several advantages
such as shortening the burn-in, accelerating the mixing of the Markov
chain at the stationary period, and simplifying tuning \citep{dahlin2015accelerating}.

The key contributions of this paper are three-fold. First, we provide
users with ample of new efficient algorithms for simulation-based
inferences, demonstrating its ease of use in practical problems. Second,
advanced machine learning simulation-based inference algorithms are not only attractive in theory, but we show
them also have good numerical performance in practice. In particular,
they are applied in both simple toy problems and a challenging scientific
problem and they give impressive results. Third, we provide a general framework
for interested users not only they can use but also they can develop new algorithms
based on smoothing instead of filtering. Consequently, as moments
are by-products of iterated smoothing, advanced methods such as proximal
optimization simulation-based inference could be implemented in our framework. 

The paper is organized as follows. In Section~\ref{sec:background},
we introduce some backgrounds of simulation-based inferences we investigate. 
In Section~\ref{sec:methods}, we describe the fixed lag smoothing algorithm in the context of partially observed Markov
models, which are later extended to different plug-and-play methodologies. We also give a brief description on advantages and disadvantages
of each methodology. Section \ref{sec:examples} presents
two toy problems, showing substantial gains for our new methods over current alternatives, while 
Section \ref{sec:EpidemicModel} illustrates the potential applications of 
some efficient methods in a challenging inference problem of fitting a malaria
transmission model to time series data. We conclude in Section
\ref{sec:conclusion} with the suggesting of the future works to
be extended. More experiments and different functionalities can also
be found in the package repository.

\section{Background of models and simulation-based inferences}\label{sec:background}
Let $\{X(t),\ t\in\mathbb{T}\}$ be a Markov process with $X(t)$
taking values in a measurable space $\mathcal{X}$ with a finite subset
$t_{1}<t_{2}<\cdots<t_{N}$ at which $X(t)$ is observed, together
with some initial time $t_{0}<t_{1}$. We write $X_{0:N}=(X_{0},\ \ldots,\ X_{N})=(X(t_{0}),\ \ldots,\ X(t_{N}))$
and hereafter for any generic sequence $\{X_{n}\}$, we shall use
$X_{i:j}$ to denote $(X_{i},\ X_{i+1},\ \ldots,\ X_{j})$. The distribution
of $X_{0:N}$ is characterized by the initial density $X_{0}\sim\mu(x_{0};\theta)$
and the conditional density of $X_{n}$ given $X_{n-1}$, written as
$f_{n}(x_{n}|x_{n-1};\theta)$ for $1\leq n\leq N$ where $\theta$
is an unknown parameter in $\mathbb{R}^{d}$. The process $\{X_{n}\}$
takes values in a measurable space $\mathcal{Y}$, being only observed
through another process $\{Y_{n},n=1,\dots,N\}$ . The observations
are assumed to be conditionally independent given $\{X_{n}\}$. Their
probability density is of the form 
\[
p_{Y_{n}|Y_{1:n-1},\ X_{0:n}}(y_{n}|y_{1:n-1},\ x_{0:n};\theta)=g_{n}(y_{n}|x_{n};\theta),
\]
for $1\leq n\leq N$. It is assumed that $X_{0:N},\mathrm{Y}_{1:N}$
have a joint density $p_{X_{0:N},Y_{1:N}}(x_{0:N},y_{1:N};\theta)$
on $\mathcal{X}^{N+1}\times\mathcal{Y}^{N}$. The data are a sequence
of observations $y_{1:N}^{*}=(y_{1}^{*},\ \ldots,\ y_{N}^{*})\in\mathcal{Y}^{N}$,
considered as fixed and we write the $\log$ likelihood function of
the data for the POMP model as 
\begin{eqnarray*}
\ell(\theta) & = & \log p_{Y_{1:N}}(y_{1:N}^{*};\theta)\\
 & = & \log\int\mu(x_{0};\theta)\prod_{n=1}^{N}f_{n}(x_{n}|x_{n-1};\theta)\ g_{n}(y_{n}^{*}|x_{n};\theta)\ dx_{0:N}.
\end{eqnarray*}
A practical limitation for those models is that it is difficult or even impossible
to compute the log-likelihood and hence to compute the MLE in closed
form. Therefore, MLE process often uses first order
stochastic approximation \citep{kushner2012stochastic}, which involves a Monte
Carlo approximation to a difference equation, 
\[
\theta_{m}=\theta_{m-1}+\gamma_{m}\nabla\ell(\theta_{m-1}),
\]
where $\theta_{0}\in\Theta$ is an arbitrary initial estimate and
$\{\gamma_{m}\}_{m\geq1}$ is a sequence of step sizes with ${\sum_{m\geq1}\gamma_{m}=\infty}$
and ${\sum_{m\geq1}\gamma_{m}^{2}<\infty}$. The algorithm converges
to a local maximum of $\ell(\theta)$ under regularity conditions.
The term $\nabla\ell(\theta)$, also called the score function, is shorthand
for the $\mathbb{R}^{d}$-valued vector of partial derivatives, 
\[
\nabla\ell(\theta)=\frac{\partial\ell(\theta)}{\partial\theta}.
\]
Sequential Monte Carlo (SMC) approaches have previously been developed
to estimate the score function \citep{poyiadjis2009sequential,nemeth2016particleB,dahlin2015particle}.
However under plug-and-play setting, which does not requires the ability
to evaluate transition densities and their derivatives, these approaches
are not applicable. Consider a parametric model consisting of a density
$p_{Y}(y;\theta)$ with the log-likelihood of the data $y^{*}\in\mathcal{Y}$
given by $\ell(\theta)=\log p_{Y}(y^{*};\theta)$. A stochastically
perturbed model corresponding to a pair of random variables $(\breve{\Theta},\breve{Y})$
having a joint probability density on $\mathbb{R}^{d}\times\mathcal{Y}$
can be defined as 
\[
p_{\breve{\Theta},\breve{Y}}(\breve{\vartheta},\ y;\theta,\ \tau)=\tau^{-d}\kappa\left\{ \tau^{-1}(\breve{\vartheta}-\theta)\right\} p_{Y}(y;\breve{\vartheta}).
\]
Suppose some regularity conditions, \citep{doucet2013derivative}
showed that 
\begin{equation}
\left|\tau^{-2}\Sigma^{-1}\breve{\mathbb{E}}\left(\breve{\Theta}-\theta\left|\breve{Y}=y^{*}\right.\right)-\nabla\ell\left(\theta\right)\right|<C\tau^{2}.\label{eq:4.1}
\end{equation}
These approximations are useful for latent variable models, where
the log likelihood of the model consists of marginalizing over a latent
variable, $X$, 
\[
\ell(\theta)=\log\int{p_{{X},{Y}}(x,y^{*};\theta)\,dx}.
\]
In latent variable models, the expectation in equation \eqref{eq:4.1} can be approximated
by Monte Carlo importance sampling, as proposed by \citet{ionides11}
and \citet{doucet2013derivative}. In \cite{nguyen2017second}, the POMP
model is a specific latent variable model with ${X}=X_{0:N}$ and
${Y}=Y_{1:N}$. A perturbed POMP model is defined to have a similar
construction to a perturbed latent variable model with ${\breve{X}}=\breve{X}_{0:N}$,
$\breve{Y}=\breve{Y}_{1:N}$ and $\breve{\Theta}=\breve{\Theta}_{0:N}$.
\citet{ionides11} perturbed the parameters by setting $\breve{\Theta}_{0:N}$
to be a random walk starting at $\theta$, whereas \citet{doucet2013derivative}
took $\breve{\Theta}_{0:N}$ to be independent additive white noise
perturbations of $\theta$. Taking advantage of the asymptotic developments
of \citet{doucet2013derivative} while maintaining some practical
advantages of random walk perturbations for finite computations, \cite{nguyen2017second}
show that $\nabla{\ell}\left(\theta\right)$ of the extended model
can be approximated as follows.
\begin{theorem} \label{thm1-1} (Theorem 2 of \citet{nguyen2017second}) Suppose assumptions 1, 2
and 5 of \citep{nguyen2017second}, there exists a constant $C$ independent of $\tau,\tau_{1},...\tau_{N}$
such that, 
\[
\left|\nabla\ell\left(\theta\right)-\tau^{-2}\Psi^{-1}\left\{ \tau_{0}^{-2}\breve{\mathbb{E}}\left(\breve{\Theta}_{0}-\theta|\breve{Y}_{1:N}=y_{1:N}^{*}\right)\right\} \right|<C\tau^{2},
\]
where $\Psi$ is the non-singular covariance matrix associated to
$\psi$. \end{theorem}
\citet{nguyen2017second} also presented an alternative variation on these
results which leads to a more stable Monte Carlo estimation. \begin{theorem}
\label{thm3}(Theorem 3 of \citet{nguyen2017second}) Suppose assumptions
1,2, 5 as in \citet{nguyen2017second}. In addition, assume that $\tau_{n}=O(\tau^{2})$
for all $n=1\ldots N$, the following hold true, 
\begin{equation}
\left|\nabla\ell\left(\theta\right)-\frac{1}{N+1}\tau^{-2}\tau_{0}^{-2}\Psi^{-1}\sum_{n=0}^{N}\left\{ \breve{\mathbb{E}}\left(\breve{\Theta}_{n}-\theta|\breve{Y}_{1:N}=y_{1:N}^{*}\right)\right\} \right|=O(\tau^{2}).
\end{equation}
\end{theorem}
These theorems are useful for our investigated simulation-based inference approaches
because we can approximate the gradient of the likelihood of the extended
model to the second order of $\tau$ using either particle filtering or particle smoothing, 
which fits well with our inference framework in the next section. 

\section{Methodology for POMP models}\label{sec:methods}

We are mostly interested in full-information, either Bayesian or frequentist and plug-and-play methods.
Table~\ref{tab:method} provides a list of several inference methodologies for POMP models implemented in \pkg{is2}. 

\begin{table}[t]
\begin{tabular}{l|p{0.35\linewidth}|p{0.35\linewidth}}
\multicolumn{3}{l}{\bf Implemented algorithms \rule[-2mm]{0mm}{4mm}  }\tabularnewline
\hline
&Frequentist   & Bayesian  \tabularnewline
\hline
Full information& Second-order iterated smoothing (\code{is2}, \cref{sec:is2}) \raggedright &Particle iterated filtering (\code{pmif}, \cref{sec:pmif}) \raggedright \tabularnewline
&Momentum iterated filtering (\code{Momentum-Mif}, \cref{sec:Momentum}) \raggedright & \raggedright \tabularnewline
&Accelerate iterated filtering (\code{aif}, \cref{sec:aif}), \raggedright & \raggedright \tabularnewline 
&Average iterated filtering (\code{avif}, \cref{sec:avif}) \raggedright & \raggedright \tabularnewline
\hline
\end{tabular}\label{tab:method}
\caption{
  Plug-and-play inference methods for POMP models, implemented in \pkg{is2}.
}
\label{tab:methods}
\end{table}

\subsection{The likelihood function and particle smoothing}\label{sec:pfilter}

The structure of a POMP model indicates its log likelihood $\loglik(\theta)$
can be expressed as the sum of conditional log likelihoods. Typically
$\loglik(\theta)$ has no closed form and is generally approximated
by Monte Carlo methods. In the primitive bootstrap particle filter,
the filtering distribution $p(\mathrm{X}_{t}|\mathrm{y}_{1:t})$ at
each time step $t$ is represented by $J$ weighted samples $(x_{t}^{j},\ w_{t}^{j})_{j=1}^{J}$,
where $(x_{t}^{j})_{j=1}^{J}$ are random samples of the distribution
and the weight $w_{t}^{j}$ is the density at this sample.
Initially, each weight is set to $\frac{1}{J}$ and $x_{0}^{j}$
are sampled from the prior distribution $\mu_{0}(dx_{0})$. At each
time step $t\geq1$, we draw a sample $X_{t}^{j}$ from $f_{X_{t}|X_{t-1}}(x_{t}|x_{t-1};\theta)$
and use observation $y_{t}$ to compute $w_{t}^{j}$ such that $w_{t}^{j}\propto g(y_{t}|x_{t}^{j})$
and then we normalize them to sum to 1. Now we sample $\left\{ X_{t}^{j}\right\} $
by resampling them with probabilities proportional $\left\{ w_{t}^{j}\right\} $.
The process goes on and the log likelihood approximation is estimated
by $\loglik(\theta)=\sum_{t=1}^{T}\log\big(J^{-1}\,\sum_{j=1}^{J}\!w_{t}^{j})$.
It has been shown that as $J\rightarrow\infty$, the approximation
will approach the true log likelihood (e.g. \citep{delMoral04}) and $J$
is typically chosen as a fixed large number.

Primitive bootstrap particle filter is often deteriorated quickly,
especially when there are outliers in the observations. This leads
to the \emph{particle depletion} problem when the distribution are
represented by only a few survival particles. For this reason, particle
smoothing is often preferable to particle filtering since the variance
of the smoothing distribution is always smaller than that of the filtering
distribution. However, a naive particle smoothing algorithm does not
get rid of the \emph{particle depletion problem} entirely and is known
for rather computationally expensive \citep{kitagawa1996monte,doucet2000sequential}. Many variations of the particle smoothing algorithm have been
proposed to ameliorate this problem. Some approaches may improve numerical
performance in appropriate situations \citep{cappe07} but typically
lose the plug-and-play property, except for fixed lag smoothing. Since
we are interested in plug-and-play property, which assists in investigating
of alternative models, we will focus here on fixed lag smoothing.
Specifically, rather than targeting the sequence of filtering densities,
we target the sequence of distribution $p(dx_{t}|y_{1:T})$. It is
often reasonable to assume that for sufficiently large $L$, the future state
after the $L$ time steps has little information about the current state.
This motivates an approximation, in which we only use all observations up to a time $s+L$ for some
sufficiently large $L$ when estimating the state at time $s$. Mathematically, this is equivalent to 
approximating marginally $p(dx_{t}|y_{1:T})$ by $p(dx_{t}|y_{1:t+L})$
for some lag value $L$, and then approximating $p(dx_{t}|y_{1:t+L})$
with the sequence of densities $p(x_{t}|y_{1:t+L})$ for $t=1,2,\ \ldots T-L$
and $p(x_{t}|y_{1:T})$ for $t=T-L+1,\ \ldots T$ (see e.g. \citep{polson2008practical}. This is done by representing $p(x_{t}|y_{1:t+L})$
with a set of weighted particles $\{x_{t}^{j},\ w_{t}^{j}\}_{j=1}^{J}$,
each of which is a state $x_{t}$. 

%%%%  PFILTER PSEUDOCODE
\begin{algorithm}[H]
  \caption{\textbf{Fixed lag particle smoothing}:
    \code{psmooth(\,P,\,Np{\argequals}$J$)}, where \code{P} is a \class{pomp} object with definitions for \code{rprocess}, \code{dmeasure}, \code{init.state}, \code{coef}, and \code{obs}.
    \label{alg:psmooth}
  }
  \KwIn{
    Simulator for $f_{X_n|X_{n-1}}(x_n\given x_{n-1}\giventh\theta)$;
    evaluator for $f_{Y_n|X_n}(y_n\given x_{n}\giventh\theta)$;
    simulator for $f_{X_0}(x_0\giventh\theta)$;
    parameter, $\theta$;
    data, $y^*_{1:N}$;
    number of particles, $J$.
  }
  \BlankLine
  Initialize smoothing particles:
  simulate ${X}_{0,j}^{F}\sim {f}_{{X}_{0}}\left(\mydot\giventh{\theta}\right)$ for $j$ in $\seq{1}{J}$.\;
  \For{$n$ in $\seq{1}{N}$}{
    Simulate for prediction:
    ${X}_{n,j}^{P}\sim {f}_{{X}_{n}|{X}_{n-1}}\big(\mydot|{X}_{n-1,j}^{F};{\theta}\big)$ for $j\ \text{in}\ \seq{1}{J}$. \nllabel{alg:pfilter:step1}\;
    Evaluate weights:
    $w(n,j)={f}_{{Y}_{n}|{X}_{n}}(y_{n}^{*}|{X}_{n,j}^{P}\giventh{\theta})$ for $j$ in $\seq{1}{J}$.\;
    Normalize weights:
    $\tilde{w}(n,j)= w(n,j)/\sum_{m=1}^{J}w(n,m)$.\;
    Apply resampling to select indices $k_{1:J}$ with $\prob{k_{j}=m} =\tilde{w}(n,m)$.\nllabel{alg:pfilter:systematic}\;
    resample:
    set ${X}_{n,j}^{F}={X}_{n,k_{j}}^{P}$ for $j$ in $\seq{1}{J}$. \nllabel{alg:pfilter:step2} \;
    for $n$ in $1:N-L$, let $a_{1}(n,k_{j})=j$, $a{}_{l+1}(n,j)=a_{1}(n-l,a_{l}(n,j))$ for $j$ in $1:J$, $l$ in $1:L-1$\;
    Compute conditional log likelihood: $\loglikMC_{n|1:n-1}=\log\big(J^{-1}\,\sum_{m=1}^{J}\!w(n+L,m)\big)$ for $n$ in $1:N-L$ and $\loglikMC_{n|1:n-1}=\log\big(J^{-1}\,\sum_{m=1}^{J}\!w(N,m)\big)$ for $n$ in $N-L+1:N$\;
  }
  \KwOut{
    Log likelihood estimate, $\loglikMC(\theta)=\sum_{n=1}^N\loglikMC_{n|1:n-1}$;
    smoothing sample, $X^F_{n+L,1:J}$, for $n$ in $\seq{1}{N-L}$ and $X^F_{N,1:J}$, for $n$ in $\seq{N-L+1}{N}$.
  }
  \KwCplx{$\bigO{J}$}
\end{algorithm}

In the fixed lag smoothing, $x_{t}^{j}$ is obtained by tracing back the lineage of the particle $x_{t+L}^{j}$. Define 
$a_{1}(x_s,k_{j})=j$, $a{}_{l+1}(x_s,j)=a_{1}(x_{s-l},a_{l}(x_s,j))$
for all $t\leq s<t+L$, then $x_{t}^{j}$ is $x_{t+L}^{a_{L}(x_{t+L},j)}$.

The algorithm of \citep{kitagawa1996monte} can be used to approximate $p(x_{t}|y_{1:t+L})$
for a suitable value $L$. Choosing an optimal $L$ might be difficult but in general
for any given lag, smoothing is often a better choice compared to filtering. Another advantage
of a fixed lag smoothing is that the computational complexity is the
same as that of the particle filter. We present the fixed lag smoothing
algorithm in Algorithm~\ref{alg:psmooth}.

%%% ITERATED FILTERING
\subsection{Second-order iterated smoothing}\label{sec:is2}

Iterated smoothing, a variant of iterated filtering algorithm, enables
inference using simulation-based sequential Monte Carlo smoothing. The
basic form of iterated smoothing, proposed by \cite{doucet2013derivative},
was found to have favorable theoretical properties such as being more stable
than iterated filtering and producing tighter moment approximations. The innovations
of \cite{doucet2013derivative} are perturbing parameters by white
noise and basing on a sequential Monte Carlo solution to the smoothing
problem. However, it is an open problem whether the original iterated
smoothing approach is competitive in practice.  \cite{nguyen2017second} modified
the theory developed by \cite{doucet2013derivative} in two key regards. 

First, random walk parameter perturbations can be used in place of
the white noise perturbations of the original iterated smoothing while
preserving much of the theoretical support provided by \cite{doucet2013derivative}.
Perturbation in this way has some positive effects:
\begin{enumerate}
\item It inherits second-order convergence properties by using an approximation
of the observed information matrix.
\item It combats computational expense by canceling out some computationally
demanding covariance terms.
\item Plug-and-play property, inherited from the sequential Monte Carlo
smoothing, is preserved. 
\end{enumerate}
Second, \cite{nguyen2017second} modified to construct an algorithm which carries
out smoothing, in particular fixed lag smoothing. This Taylor-series
approach remains competitive with other state
of the art inference approaches both in theory and in practice. 

By default, iterated smoothing carries out the fixed lag smoothing
with a specified number of lag $l$ and an option for choosing either random
walk noise or white noise perturbation. A basic version of a second-order iterated smoothing algorithm 
using random walk noise is available via the \code{method=\textquotedbl is2\textquotedbl}. The second order
iterated smoothing algorithm \code{is2} of \citet{nguyen2017second} has shown promises
in advancing development of simulation-based inferences. In all iterated
smoothing methods, by analogy with simulated annealing, it has shown
that in the limit as the perturbation intensity approaches zero, the
algorithm will reach the maximum likelihood solution. This perturbation
process is gradually reduced according to a prescribed cooling schedule
in order to allow the likelihood function to converge.

For some time series models, a few unknown initial conditions, called initial
value parameters (IVP), are estimated as parameters. As the length of the time series increases, the estimates are not
accurate since only early time points have information of the initial
state \citep{king2015statistical}. Therefore for these parameters, we only perturb them at time zero. In general, the perturbation distribution is supplied by the user
either with each component, $[\theta]_{i}$, of the parameter vector
being perturbed independently or simultaneously. By default, the perturbations
on the parameters in \cref{alg:mif:init:perturb,alg:mif:perturb}
of \cref{alg:mif} is a normal distribution and is perturbed independently.
In addition, it is always possible to transform the parameters and
choose the perturbation distribution of interest to simplify the algorithm. 
% MIF PSEUDOCODE
\begin{algorithm}[H]
  \caption{
    \textbf{Second-order iterated smoothing}:
    \texttt{is2(P, start{\argequals}$\theta_0$, Nis{\argequals}$M$, Np{\argequals}$J$, rw.sd{\argequals}$\sigma_{1:p}$, lag{\argequals}$L$, var.factor{\argequals}$C$, cooling.factor{\argequals}$a$)},
    where \code{P} is a \class{pomp} object with defined \code{rprocess}, \code{dmeasure}, \code{init.state}, and \code{obs} components.
    \label{alg:mif}
    }
  \KwIn{
    Starting parameter, $\theta_0$;
    simulator for $f_{X_0}(x_0\giventh\theta)$;
    simulator for $f_{X_n|X_{n-1}}(x_n\given x_{n-1}\giventh\theta)$;
    evaluator for $f_{Y_n|X_n}(y_n\given x_{n}\giventh\theta)$;
    data, $y^*_{1:N}$;
    labels, $I\subset\{1,\dots,p\}$, designating IVPs;
    fixed lag, $L$, for estimating IVPs;
    number of particles, $J$,
    number of iterations, $M$;
    cooling rate, $0<a<1$;
    perturbation scales, $\sigma_{1:p}$;
    initial scale multiplier, $C>0$.
  }
  \BlankLine
  \For {$m$ in $\seq{1}{M}$}{
    $\big[\Theta_{0,j}^{F}\big]_{i}\sim\mathcal{N}\big(\big[\theta_{m-1}\big]_{i},(c^{m-1}\sigma_{i})^{2}\big)$ for $j$ in $1:J$, $i$ in $1:d$ \;
    initialize states: simulate $X_{0,j}^{F}\sim \mu\big({x}_{0};{\Theta_{0,j}^{F}}\big)$ for {$j$ in $1:J$}\;
    \For {$n$ \textrm{in} $1\myto N$}{
      Filtering: \nllabel{alg:is2:filtering}
      $\big[\Theta_{n,j}^{P}\big]_{i}\sim\mathcal{N}\big(\big[\Theta^F_{n-1,j}\big]_{i},(c^{m-1}\sigma_{i})^{2}\big)$ for $i\notin I$, $j$ in $1:J$\;
      Simulate particles: \nllabel{alg:mif:sim}
      ${X}_{n,j}^{P}\sim{f}_n\big({x}_{n}|{X}_{n-1,j}^{F};{\Theta_{n,j}^{P}}\big)$ for $j$ in $1:J$\label{alg:mif:sim}\;
      Evaluate weights: \nllabel{alg:mif:weights}
      $w(n,j)=g_n(y_{n}^{*}|X_{n,j}^{P};\Theta_{n,j}^{P})$ for $j$ in $1:J$\;
      Normalize weights: \nllabel{alg:mif:normalize}
      $\breve{w}(n,j)=w(n,j)/\sum_{u=1}^{J}w(n,u)$\;
      Apply resampling to select indices $k_{1:J}$ with $P\left\{ k_{u}=j\right\} =\breve{w}\left(n,j\right)$\nllabel{alg:mif:syst}\;
      resample particles: \nllabel{alg:mif:resample}
      $X_{n,j}^{F}=X_{n,k_{j}}^{P}$ and $\Theta_{n,j}^{F}=\Theta_{n,k_{j}}^{P}$ for $j$ in $1:J$\;
      fixed-lag smoothing: 
      for $n$ in $1:N-L$, let $a_{1}(n,k_{j})=j$, $a{}_{l+1}(n,j)=a_{1}(n-l,a_{l}(n,j))$ for $j$ in $1:J$, $l$ in $1:L-1$\;
      smoothed mean: $\bar{\theta}_{n-L}^{L}=\sum_{j=1}^{J}\breve{w}(n,j)\Theta_{n-L,a_{L}(n,j)}^{P}$ if $n>L$\;
      variance:${V}_{n-L,n-L}^{m}=\sum_{j}\breve{w}(n,j)\big(\Theta_{n-L,a_{L}(n,j)}^{P}-\bar{\theta}_{n-L}^{L}\big)\big(\Theta_{n-L,a_{L}(n,j)}^{P}-\bar{\theta}_{n-L}^{L}\big)^{\top}$ if $n>L$\;
    }
    for $n$ in $N-L+1:N$ mean: $\bar{\theta}_{N+l-L}^{L}=\sum_{j=1}^{J}\breve{w}(N,j)\Theta_{N+l-L,a_{L-l}(N,j)}^{P}$ for $l$ in $1:L$\;
    $[\theta_m]_i=[\theta_{m-1}]_i+V_{1}^{i}\sum_{n=1}^{N}\left(V_{n}^{i}\right)^{-1}\left(\bar{\theta}_{n}^{i}-\bar{\theta}_{n-1}^{i}\right)$ for $i \not\in I$.\;
    ${V}_{N+l-L,N+l-L}^{m}=\sum_{j}\breve{w}(N,j)\big(\Theta_{N+l-L,a_{L-l}(N,j)}^{P}-\bar{\theta}_{N+l-L}^{L}\big)$\newline${}$\hspace{24mm}$\big(\Theta_{N+l-L,a_{L-l}(N,j)}^{P}-\bar{\theta}_{N+l-L}^{L}\big)^{\top}$ for $l$ in $1:L$\;
    update: $S_{m}=c^{-2(m-1)}\Psi^{-1}\sum_{n=1}^N\big[\left(\bar{\theta}_{n}^{L}-\theta_{m-1}\right)\big]$\;
    $I_{m}=-c^{-4(m-1)}\Psi^{-1}\left[\sum_{n=1}^{N}\left(V_{n,n}^{m}/(N+1)-c^{2(m-1)}\Psi\right)\right]\Psi^{-1}$\label{alg:pd}\;
    update parameters: $\theta_{m}=\theta_{m-1}+I_{m}^{-1}S_{m}$\;
    update IVP parameters: $\big[\theta_{m}\big]_i=\frac{1}{J}\sum_{j=1}^J\big[\Theta^F_{L,j}\big]_i$ for $i \in I$\;
  }
  \KwOut{Monte~Carlo maximum likelihood estimate, $\theta_M$.}
  \KwCplx{$\bigO{J M}$}
\end{algorithm}

This algorithm is especially useful for parameter estimation in state-space
models, in which the focus is on continuous space, discrete time and
the latent process can be simulated and the conditional density of
the observation given the state can be evaluated point-wise up to
a multiplicative constant. The parameters are often added to the state
space and this modified state-space model has been inferred using
sequential Monte Carlo filtering/smoothing. In the literature, there has been
a number of previous proposed approaches to estimate the parameter such
as \citep{kitagawa98,janeliu01,wan00} but second-order iterated smoothing
is distinguished by having an optimal theoretical convergence rate to
the maximum likelihood estimate  and is fully supported in practice
as well. 

%%% MOMENTUM ITERATED FILTERING
\subsection{Momentum iterated filtering}\label{sec:Momentum}
The underlying theoretical foundation of original iterated filtering \citep{ionides06} is 
stochastic gradient descent (SGD). It is generally well known that SGD would easily get 
stuck in ravines where the curvature surfaces are steeper in some dimensions than the others. 
It would oscillate across the slopes of the ravine and slowly progress to the local optimum, 
thus slow down the convergence rate. Some motivations to apply momentum method to iterated 
filtering rather than using the original iterated filtering consist of:
\begin{itemize}
 \item To accelerate the convergence rate, it accumulates a velocity vector in directions of 
persistent increase in the log likelihood function across iterations \citep{sutskever2013importance}. 
Specifically, if the dimensions of the momentum term and the gradient are in the same directions, 
movement in these directions will be accelerated. On the other hand, if the dimensions of the 
momentum term and the gradient are in the opposite directions, the momentum term will lend a 
helping hand to dampen the oscillation in these directions. It can be shown that momentum approach 
requires fewer iterations to get to the identical degree of precision than gradient descent 
approach  \citep{polyak1964gradient}.
 \item Instead of re-weighting the update throughout each eigen-direction proportional to the 
inverse of the associated curvature, momentum method adapts the learning rate by accumulating 
changes along the updated path. For high dimensional data, it is notoriously admitted that 
explicit computation of curvature matrix is expensive or even infeasible. Accumulating this 
information along the way appears to possess a certain computational advantage, especially 
for sparse data regime.
 \item Another crucial merit of momentum method is the capability to avoid getting trapped in 
its numerous suboptimal local maxima or saddle points. \citep{dauphin2014identifying} postulated that 
the difficulty arises indeed not from local maxima but from saddle points, i.e. points where 
one dimension slants up and another inclines down. These saddle points are usually surrounded 
by a plateau of the same error, which makes it notoriously difficult for SGD to get away from, 
as the gradient is near zero in all dimensions. It is vital to examine to what degree, momentum 
method escapes such saddle points, particularly, in the high-dimensional setting in which the 
directions of avoiding points may be few.
\end{itemize}
\begin{algorithm}[h]
  \caption{
    \textbf{Momentum iterated filtering}:
    \texttt{mif-Momentum(P, start{\argequals}$\theta_0$, Nmif{\argequals}$M$, Np{\argequals}$J$, rw.sd{\argequals}$\sigma_{1:p}$, ic.lag{\argequals}$L$, var.factor{\argequals}$C$, cooling.factor{\argequals}$a$)},
     where \code{P} is a \class{pomp} object with defined \code{rprocess}, \code{dmeasure}, \code{init.state}, and \code{obs} components.
    \label{alg:Momentum}
    }
  \KwIn{
    Starting parameter, $\theta_0$;
    simulator for $f_{X_0}(x_0\giventh\theta)$;
    simulator for $f_{X_n|X_{n-1}}(x_n\given x_{n-1}\giventh\theta)$;
    evaluator for $f_{Y_n|X_n}(y_n\given x_{n}\giventh\theta)$;
    data, $y^*_{1:N}$;
    labels, $I\subset\{1,\dots,p\}$, designating IVPs;
    fixed lag, $L$, for estimating IVPs;
    number of particles, $J$,
    number of iterations, $M$;
    cooling rate, $0<a<1$;
    perturbation scales, $\sigma_{1:p}$;
    initial scale multiplier, $C>0$;
    momentum term $\gamma$.
  }
  \BlankLine
  \For {$m$ in $\seq{1}{M}$}{
    Initialize parameters: \nllabel{alg:mif:init:perturb}
    $[\Theta^F_{0,j}]_i \sim \normal\left([\theta_0]_i, (C a^{m-1} \sigma_i)^2\right)$ for $i$ in $\seq{1}{p}$, $j$ in $1\myto J$.\;
    Initialize states: \nllabel{alg:mif:initstates}
    simulate $X_{0,j}^F \sim f_{X_0}\big(\mydot;{\Theta^F_{0,j}}\big)$ for $j$ \textrm{in} $\seq{1}{J}$.\;
    Initialize filter mean for parameters:
    $\bar \theta_0=\theta_0$.\;
    \For {$n$ \textrm{in} $1\myto N$}{
      Perturb parameters: \nllabel{alg:mif:perturb}
      $\big[\Theta^P_{n,j}\big]_i\sim\normal\left(\big[\Theta^F_{n-1,j}\big]_i,(a^{m-1} \sigma_i)^2\right)$ for $i \not\in I$, $j$ in $1\myto J$.\;
      Simulate prediction particles: \nllabel{alg:mif:sim}
      ${X}_{n,j}^{P}\sim {f}_{{X}_{n}|{X}_{n-1}}\big(\mydot|{X}_{n-1,j}^{F}\giventh{\Theta^P_{n,j}}\big)$ for $j$ in $1\myto J$.\;
      Evaluate weights: \nllabel{alg:mif:weights}
      $w(n,j)=f_{Y_{n}|X_{n}}(y_{n}^{*}|X_{n,j}^{P}\giventh\Theta^P_{n,j})$ for $j$ in $1\myto J$.\;
      Normalize weights: \nllabel{alg:mif:normalize}
      $\tilde{w}(n,j)= w(n,j)/\sum_{u=1}^{J}w(n,u)$.\;
      Apply resampling to select indices $k_{1:J}$ with $\prob{k_{u}=j} =\tilde{w}\left(n,j\right)$.\nllabel{alg:mif:syst}\;
      resample particles: \nllabel{alg:mif:resample}
      $X_{n,j}^{F}=X_{n,k_{j}}^{P}$ and $\Theta_{n,j}^{F}=\Theta^P_{n,k_{j}}$ for $j$ in $1\myto J$.\;
      Filter mean: \nllabel{alg:mif:filtermean}
      $\big[\bar{\theta}_{n}\big]_i=\sum_{j=1}^J\tilde{w}(n,j)\big[\Theta^P_{n,j}\big]_i$ for $i \not\in I$.\;
      Prediction variance: \nllabel{alg:mif:predvar}
      $[\bar{V}_{n+1}]_i=(a^{m-1}\sigma_i)^2 + \sum_{j}\tilde{w}(n,j)\big([\Theta^F_{n,j}]_i - [\bar{\theta}_{n}]_i\big)^2$ for $i \not\in I$.\;
    }
    $\mu_0^i=[\theta_0]_i$.\;
     $\mu_m^i= \gamma\mu_{m-1}^i- V_{1}^{i}\sum_{n=1}^{N}\left(V_{n}^{i}\right)^{-1}\left(\bar{\theta}_{n}^{i}-\bar{\theta}_{n-1}^{i}\right)$ for $i \not\in I$.\;
    $[\theta_m]_i=[\theta_{m-1}]_i- \mu_m^i$.\;
  }
  \KwOut{$\theta_M$.}
  \KwCplx{$\bigO{J M}$}
\end{algorithm}

While the classical convergence theories for momentum method depend on noiseless gradient 
approximate, it is possible to extend to stochastic setting of iterated filtering. 
Unfortunately, it has been conjectured that any merit in terms of asymptotic local rate of 
convergence will be lost \citep{wiegerinck1994stochastic}, and \citet{lecun2012efficient} 
confirmed 
that in extensive experiments. However, fascination in applying momentum methods has received 
considerable attentions recently  after they diminished for a while. In simulation-based inference,  
the momentum method improves the convergence rate of SGD by adding a short term memory, 
a fraction $\gamma$ of the update vector of the previous one time step to the current update vector. 
Note that, momentum-based methods will  outperform SGD for convex objective functions in 
the early or transient phase of the optimization process where l/t is the most influential term 
though both the methods will be equally effective during the ending phase. However, it is an 
open question whether momentum-based methods yield faster rates in the non-convex setting, 
specifically when we consider the convergence criterion of second-order stationarity. Motivated 
by performance of momentum-based methods (e.g. \citep{sebastian16}), we implement momentum iterated filtering algorithm in Algorithm \ref{alg:Momentum}. We
only focus on providing potential interesting methodologies for simulation-based inference 
while its theoretical foundations will be published elsewhere.

%%% ACCELERATE ITERATED FILTERING
\subsection{Accelerate iterated filtering}\label{sec:aif}

Some motivations to accelerate iterated filtering rather than the naive
iterated filtering include: 
% AIF PSEUDOCODE
\begin{algorithm}[h]
  \caption{
    \textbf{Accelerate iterated filtering}:
    \texttt{aif(P, start{\argequals}$\theta_0$, Nmif{\argequals}$M$, Np{\argequals}$J$, rw.sd{\argequals}$\sigma_{1:p}$, ic.lag{\argequals}$L$, var.factor{\argequals}$C$, cooling.factor{\argequals}$a$)},
    where \code{P} is a \class{pomp} object with defined \code{rprocess}, \code{dmeasure}, \code{init.state}, and \code{obs} components.
    \label{alg:aif}
    }
  \KwIn{
    Starting parameter, $\theta_0=\theta^{ag}_0$, sequences $\alpha_n, \beta_n, \lambda_n, \Gamma_n$;
    simulator for $f_{X_0}(x_0\giventh\theta)$;
    simulator for $f_{X_n|X_{n-1}}(x_n\given x_{n-1}\giventh\theta)$;
    evaluator for $f_{Y_n|X_n}(y_n\given x_{n}\giventh\theta)$;
    data, $y^*_{1:N}$;
    labels, $I\subset\{1,\dots,p\}$, designating IVPs;
    fixed lag, $L$, for estimating IVPs;
    number of particles, $J$,
    number of iterations, $M$;
    cooling rate, $0<a<1$;
    perturbation scales, $\sigma_{1:p}$;
    initial scale multiplier, $C>0$.
  }
  \BlankLine
  Initialize parameters: \nllabel{alg:mif:init:perturb}
  $\theta^{md}_0=\theta_0$.\; 
  $[\Theta^F_{0,j}]_i \sim N\left([\theta^{md}_0]_i, (C a^{m-1} \sigma_i)^2\right)$ for $i$ in $1..p$, $j$ in $1... J$.\;
  Initialize states: \nllabel{alg:mif:initstates}
  simulate $X_{0,j}^F \sim f_{X_0}\big(\cdot;{\Theta^F_{0,j}}\big)$ for $j$ \textrm{in} $1..J$.\;
  \For {$m$ \textrm{in} $\seq{1}{M}$}{
    $\theta^{md}_m= (1-\alpha_m)\theta^{ag}_{m-1}+\alpha_{m}\theta_{m-1}$.\;
    \For {$n$ \textrm{in} $1... N$}{
      Perturb parameters: \nllabel{alg:mif:perturb}
      $\big[\Theta_{n,j}^{P}\big]_{i}\sim\mathcal{N}\big(\big[\theta^{md}_m\big]_{i},(c^{m-1}\sigma_{i})^{2}\big)$ for $i\notin I$, $j$ in $1:J$.\;
      Simulate prediction particles: \nllabel{alg:mif:sim}
      ${X}_{n,j}^{P}\sim{f}_n\big({x}_{n}|{X}_{n-1,j}^{F};{\Theta_{n,j}^{P}}\big)$ for $j$ in $1:J$.\;
      Evaluate weights: \nllabel{alg:mif:weights}
      $w(n,j)=g_n(y_{n}^{*}|X_{n,j}^{P};\Theta_{n,j}^{P})$ for $j$ in $1:J$.\;
      Normalize weights: \nllabel{alg:mif:normalize}
      $\breve{w}(n,j)=w(n,j)/\sum_{u=1}^{J}w(n,u)$.\;
      Apply resampling to select indices $k_{1:J}$ with $P\left\{ k_{u}=j\right\} =\breve{w}\left(n,j\right)$.\nllabel{alg:mif:syst}\;
      resample particles: \nllabel{alg:mif:resample}
      $X_{n,j}^{F}=X_{n,k_{j}}^{P}$ and $\Theta_{n,j}^{F}=\Theta_{n,k_{j}}^{P}$ for $j$ in $1:J$.\;
      Filter mean: \nllabel{alg:mif:filtermean}
      $\bar{\theta}_{n}=\sum_{j=1}^{J}\breve{w}(n,j)\Theta_{n,j}^{F}$\;
    }
    Update Parameters: \nllabel{alg:mif:update}
    $S_{m}=c^{-2(m-1)}\Psi^{-1}\sum_{n=1}^N\big[\left(\bar{\theta}_{n}-\theta^{md}_{m}\right)\big]/(N+1)$.\;
    $\big[\theta_{m}\big]_i=\theta_{m-1}-\lambda_{m-1}\big[S_{m}\big]_{i}$ for $i \notin I$.\;
    $\big[\theta^{ag}_{m}\big]_i=\theta^{md}_{m-1}-\beta_{m-1}\big[S_{m}\big]_{i}$ for $i \notin I$.\;
    $\big[\theta_{m}\big]_i=\frac{1}{J}\sum_{j=1}^J\big[\Theta^F_{j}\big]_i$ for $i \in I$.\;
  }
  \KwOut{Monte~Carlo maximum likelihood estimate, $\theta_M$.}
  \KwCplx{$\bigO{J M}$}
\end{algorithm}

\begin{enumerate}
\item \label{whyFeature2} The naive stochastic gradient descent method
converges for a general non-convex optimization problem but it does
not achieve the optimal rate of convergence, in terms of the functional
optimality gap \citep{ghadimi2016accelerated}. 
\item \label{whyFeature1} The accelerated gradient method in \cite{nesterov2013introductory}
is optimal for solving convex optimization problems, but does not
necessarily converge for solving non-convex optimization problems. 
\item \label{whyFeature3} Modified accelerated gradient method, which can
converge in both convex and non-convex optimization, assumes unbiased
estimation of the gradient which is not satisfies for most simulation-based
inferences. 
\item \label{whyFeature4} Biased estimation of the score function from sequential
Monte Carlo filtering/smoothing and perturbation noise may slow down convergence
rate of the inference, limiting the range of applicabilities. 
\item \label{whyFeature5} Accelerate gradient approaches are typically
first-order, meaning that they require only gradient, not Hessian,
which is computationally expensive. Thus, it is preferred to second-order
approaches for any big data models since it has the same convergence
rate. 
\end{enumerate}
When targeting item \ref{whyFeature3}, one aims to find a step size
scheme which can both control the added noise and doesn't affect the
convergence rate of the accelerated gradient methods. Items \ref{whyFeature1}
and \ref{whyFeature4} deliberately look for momentum which increases
the convergence rate; in this context the momentum is implicit.
It has been shown that for some selections of step sizes and moment
schemes $\left\{ \alpha_{k}\right\} ,\left\{ \beta_{k}\right\} $,$\left\{ \lambda_{k}\right\} $
and $\left\{ \Gamma_{k}\right\} $, accelerated gradient method actually
converges with an optimal convergence rate in both convex and non-convex
case. It is in the same order of magnitude as that for the original
gradient descent method \citep{nesterov2005smooth} as long as the
perturbation noise is controlled properly. We now add a few remarks
about the extension of this approach. This approach is considered
quite general since it does not require convex surface of the log-likelihood
and the estimation could be biased, especially for the simulation-based
estimation. It could be called accelerate inexact gradient approach.

Simulation-based inference of an accelerate iterated filtering is
implemented by \code{aif} (\cref{alg:aif}). The function requires
an exponentially decaying average of the moments to increase the convergence
rate. Working on this scheme, there
is substantial improvement between the aif approaches of \citet{nguyen2017second}
and the naive approach of \citet{ionides06}.
Numerical optimization of the accelerate iterated filtering is implemented
by \code{aif}, which offers a choice of the filtering method \citet{ionides11}
or smoothing method \citep{nguyen2017second}.

%%% AVERAGE ITERATED FILTERING
\subsection{Average iterated filtering}\label{sec:avif}
For class of iterated filtering approaches, under rather general conditions,
averaging over the output of each filter will result in optimal convergence
rates \citep{ruppert1988efficient, polyak1992acceleration,kushner1993stochastic, kushner2010stochastic}. 
The parameter estimation
corresponding to this averaging is defined as:
$${\displaystyle \overline{\Theta}_{k}=\frac{1}{k+1}\sum_{j=0}^{k}\Theta_{j}}.$$ 
Averaging iterated filtering has typically been motivated by two time
scales optimization process, in which the slower scale will be sped
up by the faster scale. For stochastic approximation, it is advisable
to decrease the step size slower to improve stability for finite samples
\citep{spall2005introduction}. The common practice often chooses the sequence of
step size in order of $O(n^{-\beta})$ where $\beta$ is closer to
$1/2$. This, however, makes the estimate sequence converge rather slow.
The second time scale will step in and improve the overall convergence
rate without affecting the stability of the algorithm. The key theoretical
insight behind averaging algorithms is that it is a simplest form
of two time scale algorithm and it can be implemented on-line without
remembering the previous estimation. No additional constraints is imposed
on the algorithm while the computation of averaging is simple, making
it a very attractive approach. 

One interesting feature of averaging is that it can be applied for
biased sequences (e.g. sequences generated by the Kiefer-Wolfowitz
or SPSA algorithms \citep{spall2005introduction}). In simulation-based inference, by perturbing
the sequence and using sequential Monte Carlo filter to estimate,
the output of each filter is biased, which is especially suitable in the averaging
framework. However, as shown by \citet{dippon1997weighted}, in general
averaging does not yield asymptotically optimal convergence rate.
For example, averaging across transients will only make things worse
\citep{spall2005introduction}. Therefore it is recommended to average only after some
warm start: 
$${\displaystyle \overline{\Theta}_{k}=\frac{1}{k+1-k_{Start}}\sum_{j=k_{Start}}^{k}\Theta_{j}}.$$ 
It is straightforward to apply averaging to any of the available estimators.
We only focus on average iterated filtering here as it improves estimation
in the naive iterated filtering. The average approximation in \cref{alg:avif}
by default is from all filtering output but it is optional to specified
when to start averaging. The average iterated filtering algorithm
\code{avif} is averaged
stochastic gradient descent implementation of sequential Monte Carlo
filtering (SMC) \citep{lindstrom2013tuned}. In the same style
as iterated filtering, we assume a Gaussian random walk in parameter
space. The package also supports different choices of proposal distribution. 

% AVERAGEIF PSEUDOCODE
\begin{algorithm}[h]
  \caption{
    \textbf{Average iterated filtering}:
    \texttt{avif(P, start{\argequals}$\theta_0$, Nmif{\argequals}$M$, Np{\argequals}$J$, rw.sd{\argequals}$\sigma_{1:p}$, ic.lag{\argequals}$L$, var.factor{\argequals}$C$, cooling.factor{\argequals}$a$)},
    where \code{P} is a \class{pomp} object with defined \code{rprocess}, \code{dmeasure}, \code{init.state}, and \code{obs} components.
    \label{alg:avif}
    }
  \KwIn{
    Starting parameter, $\theta_0$;
    simulator for $f_{X_0}(x_0\giventh\theta)$;
    simulator for $f_{X_n|X_{n-1}}(x_n\given x_{n-1}\giventh\theta)$;
    evaluator for $f_{Y_n|X_n}(y_n\given x_{n}\giventh\theta)$;
    data, $y^*_{1:N}$;
    labels, $I\subset\{1,\dots,p\}$, designating IVPs;
    fixed lag, $L$, for estimating IVPs;
    number of particles, $J$,
    number of iterations, $M$;
    cooling rate, $0<a<1$;
    perturbation scales, $\sigma_{1:p}$;
    initial scale multiplier, $C>0$.
  }
  \BlankLine
  \For {$m$ in $\seq{1}{M}$}{
    Initialize parameters: \nllabel{alg:mif:init:perturb}
    $[\Theta^F_{0,j}]_i \sim \normal\left([\theta_0]_i, (C a^{m-1} \sigma_i)^2\right)$ for $i$ in $\seq{1}{p}$, $j$ in $1\myto J$.\;
    Initialize states: \nllabel{alg:mif:initstates}
    simulate $X_{0,j}^F \sim f_{X_0}\big(\mydot;{\Theta^F_{0,j}}\big)$ for $j$ \textrm{in} $\seq{1}{J}$.\;
    Initialize filter mean for parameters:
    $\bar \theta_0=\theta_0$.\;
    \For {$n$ \textrm{in} $1\myto N$}{
      Perturb parameters: \nllabel{alg:mif:perturb}
      $\big[\Theta^P_{n,j}\big]_i\sim\normal\left(\big[\Theta^F_{n-1,j}\big]_i,(a^{m-1} \sigma_i)^2\right)$ for $i \not\in I$, $j$ in $1\myto J$.\;
      Simulate prediction particles: \nllabel{alg:mif:sim}
      ${X}_{n,j}^{P}\sim {f}_{{X}_{n}|{X}_{n-1}}\big(\mydot|{X}_{n-1,j}^{F}\giventh{\Theta^P_{n,j}}\big)$ for $j$ in $1\myto J$.\;
      Evaluate weights: \nllabel{alg:mif:weights}
      $w(n,j)=f_{Y_{n}|X_{n}}(y_{n}^{*}|X_{n,j}^{P}\giventh\Theta^P_{n,j})$ for $j$ in $1\myto J$.\;
      Normalize weights: \nllabel{alg:mif:normalize}
      $\tilde{w}(n,j)= w(n,j)/\sum_{u=1}^{J}w(n,u)$.\;
      Apply resampling to select indices $k_{1:J}$ with $\prob{k_{u}=j} =\tilde{w}\left(n,j\right)$.\nllabel{alg:mif:syst}\;
      resample particles: \nllabel{alg:mif:resample}
      $X_{n,j}^{F}=X_{n,k_{j}}^{P}$ and $\Theta_{n,j}^{F}=\Theta^P_{n,k_{j}}$ for $j$ in $1\myto J$.\;
      Filter mean: \nllabel{alg:mif:filtermean}
      $\big[\bar{\theta}_{n}\big]_i=\sum_{j=1}^J\tilde{w}(n,j)\big[\Theta^P_{n,j}\big]_i$ for $i \not\in I$.\;
    }
    Update non-IVP parameters: \nllabel{alg:mif:update}
    $[\theta_m]_i=(N-k_{start})^{-1}\sum_{n=k_{start}+1}^{N}\left(\bar{\theta}_{n}^{i}-\bar{\theta}_{n-1}^{i}\right)$ for $i \not\in I$.\;
    Update IVPs: \nllabel{alg:mif:updateivps}
    $[\theta_m]_i=\frac{1}{J}\sum_{j}\big[\Theta_{L,j}^F\big]_i$ for $i \in I$.\;
  }
  \KwOut{Monte~Carlo maximum likelihood estimate, $\theta_M$.}
  \KwCplx{$\bigO{J M}$}
\end{algorithm}

\subsection{Particle iterated filtering}\label{sec:pmif}

In Bayesian inference, we are interested in sampling from the parameter
posterior distribution, 
$$p({\displaystyle \theta|y_{1:T})=\frac{p(y_{1:T}|\theta)p(\theta)}{p(y_{1:T})}},$$
where $p(\theta)$ denotes the prior distribution of the parameter
and $p(y_{1:T}|\theta)=\prod_{t=1}^{T}p_{\theta}(y_{t}|y_{1:t-1})$
denotes the likelihood function.

%% PIF PSEUDOCODE
\begin{algorithm}[h]
\caption{
  \textbf{Particle iterated filtering}:
  \texttt{pmif(P, start{\argequals}$\theta_0$, Nmcmc{\argequals}$M$, Np{\argequals}$J$, proposal{\argequals}$q$)},
   where \code{P} is a \class{pomp} object with defined methods for
  \code{rprocess}, \code{dmeasure}, \code{init.state}, \code{dprior}, and \code{obs}.  The supplied \code{proposal} samples from a symmetric, but otherwise arbitrary, MCMC proposal distribution, $q(\theta^P\given\theta)$.
  \label{alg:pmif}
}
\KwIn{
  Starting parameter, $\theta_0$;
  simulator for $f_{X_0}(x_0\given\theta)$;
  simulator for $f_{X_n|X_{n-1}}(x_n\given x_{n-1}\giventh\theta)$;
  evaluator for $f_{Y_n|X_n}(y_n\given x_{n}\giventh\theta)$;
  simulator for $q(\theta^P\given\theta)$;
  data, $y^*_{1:N}$;
  number of particles, $J$;
  number of filtering operations, $M$;
  perturbation scales, $\sigma_{1:p}$;
  evaluator for prior, $f_{\Theta}(\theta)$.
}
\BlankLine
Initialization: compute $\loglikMC(\theta_0)$ using \code{pfilter} with $J$ particles.\;
\For {$m$ in $\seq{1}{M}$}{
  Approximate, $\nabla\loglikMC(\theta_{m-1})$ through mif algorithm.\;
  Draw a parameter proposal, $\theta^P_m$, from the proposal distribution:
  $\Theta^P_m \sim q\left(\mydot\given\theta_{m-1}+\nabla\loglikMC(\theta_{m-1})\right)$.\;
  Compute $\loglikMC(\theta^P_m)$ using \code{pfilter} with $J$ particles.\;
  Generate $U\sim\mathrm{Uniform}(0,1).$\;
  Set $\big(\theta_m,\loglikMC(\theta_m)\big)=\begin{cases}
  \big(\theta^P_m,\loglikMC(\theta^P_m)\big), &\text{if } U<\displaystyle\frac{f_\Theta(\theta^P_m)\exp(\loglikMC(\theta^P_m))}{f_\Theta(\theta_{m-1})\exp(\loglikMC(\theta_{m-1}))},\\
  \big(\theta_{m-1},\loglikMC(\theta_{m-1})\big),&\text{otherwise.}
  \end{cases}$\;
}
\KwOut{
  Samples, $\theta_{1:M}$, representing the posterior distribution, $f_{\Theta|Y_{1:N}}(\theta\given y_{1:N}^*)$.
}
\KwCplx{$\bigO{J M}$}
\end{algorithm}
%%%%%%%%%%%%%%%%%%%%%%%%%%%%%%%%%%%%%%%%%%%%%%%%%%%%%%%%%%%%

Particle Markov chain Monte Carlo (PMCMC) algorithms \citep{andrieu10}
enable routine full-information plug-and-play Bayesian inference for
POMP models. The basic PMCMC algorithm, called particle
marginal Metropolis-Hastings (PMMH), relies on the fact that the unbiased
likelihood estimate from SMC can be used to move the current parameter
to the desired target posterior distribution \citep{andrieu09}. It
has been shown that using additional information such as the gradient
of the log likelihood can improve the proposal distribution, resulting
in a better mixing algorithm. Thus, \citep{dahlin2015accelerating, nemeth2016particle}
proposed a particle version of Langevin algorithm. In the context
of plug-and-play algorithm, we propose a particle version of iterated
filtering algorithm where gradients of the log likelihood as the by-product
of iterated filtering are used to improve proposal distribution. This
particle iterated filtering (PIF) is implemented in \code{pmif},
as described in \cref{alg:pmif}. 

Some considerations follow, closely related to the considerations
for PMMH.
\begin{enumerate}
\item For a sufficiently small perturbation, the bias of estimation of the
gradient can be absorbed in the acceptance/rejection step and asymptotically
the posterior distribution will stabilize around the true posterior. 
\item On account of perturbation, particle iterated filtering is a doubly
plug-and-play algorithm.
\item Potentially, particle iterated filtering could be preferable to particle
Markov chain Monte Carlo in some situations. It has been argued that
information from bias estimated of the gradient of log likelihood can
increase the rather slow convergence rate and reduce the computation
complexity of PMCMC. 
\item As noted by \citet{andrieu2015convergence,andrieu2016establishing}, the mixing of the final
target distribution of the algorithm is determined by the accuracy
of the likelihood approximation. The better likelihood estimation
becomes, the better proposal will be, leading to the better mixing algorithm.
\end{enumerate}
In addition, one advantage of PIF is that the number of particles
used for SMC only increases linearly as the number of observation
increases for the same level of mixing. As a result, the computational
complexity of PIF is the same as that of PMCMC. 

\section{Model construction and data analysis: simple examples} \label{sec:examples}

To measure the performance of the new inference algorithms, we evaluate
our implemented algorithms on some benchmark examples and compare
them to the existing simulation-based approaches. We make use
of well tested and maintained code of R packages such
as \pkg{pomp} \citep{king2015statistical}. Specifically, models are coded using C
snipet declarations. New algorithms are written in
R package \pkg{is2}, which provides user-friendly interfaces in R and efficient
underlying algorithms in C.
All the simulation-based approaches mentioned above use either bootstrap particle filtering or fixed lag particle smoothing. Experiments
were carried out on a node of $32$ cores Intel Xeon E5-2680 $2.7$
Ghz with $256$ GB memory. For a fair comparison, we use the
same setup and assessment for every inference method. A public Github
repository containing scripts for reproducing our results may be found
at \url{https://github.com/nxdao2000/JSScomparisons}.

We consider the bivariate discrete time Gaussian process example and the Gompertz example from \citep{king2015statistical}.
These are two relatively simple models which we can obtain exact value of the
likelihood functions by applying Kalman filter directly or to transformed
scale. We will use likelihood computed by Kalman filter to justify
our estimation from simulation-based algorithms. Both datasets are
available from the R package \pkg{pomp} \citep{king2015statistical} by simply loading
through command \code{pompExample(ou2)} and \code{pompExample(Gompertz)}.

\subsection{Toy example: A linear, Gaussian model}

The linear model is given by the state space forms: $X_{n}|X_{n-1}=x_{n-1}\sim\mathcal{N}(\alpha x_{n-1},\sigma^{\top}\sigma),$
$Y_{n}|X_{n}=x_{n}\sim\mathcal{N}(x_{n},I_{2})$ where $\alpha$,
$\sigma$ are $2\times2$ matrices and $I_{2}$ is $2\times2$ identity
matrix. The data are simulated from the following parameters: 
\[
{\displaystyle \alpha=\left[\begin{array}{cc}
\alpha_{1} & \alpha_{2}\\
\alpha_{3} & \alpha_{4}
\end{array}\right]=\left[\begin{array}{cc}
0.8 & -0.5\\
0.3 & 0.9
\end{array}\right],\ \sigma=\left[\begin{array}{cc}
3 & 0\\
-0.5 & 2
\end{array}\right].}
\]
The number of time points $N$ is set to $100$ and initial starting
point $X_{0}=\left(-3,4\right)$. There are different specified languages for modeling such as 
BUGS \citep{spiegelhalter1996bugs}, NIMBLE \citep{de2017programming}, STAN \citep{carpenter2017stan}. 
Since we focus
on plug-and-play methods, where we only need a simulator for the transition
dynamic of pomp model, we construct our models base on C snipet of \pkg{pomp}. Interested readers are encouraged to read \citep{king2015statistical} for
the details of setting up a model. We specify the simulation process, evaluation process
and initial condition for our linear model as follows.

\begin{verbatim}
ou2.proc.sim <- function (x, t, params, delta.t, ...) {
    xi <- rnorm(n=2,mean=0,sd=1) # noise terms
    xnew <- c(
        params["alpha.1"]*x["x1"]+params["alpha.3"]*x["x2"]+
            params["sigma.1"]*xi[1],
        params["alpha.2"]*x["x1"]+params["alpha.4"]*x["x2"]+
            params["sigma.2"]*xi[1]+params["sigma.3"]*xi[2]
    )
    names(xnew) <- c("x1","x2")
    xnew
}

ou2.meas.sim <- function (x, t, params, ...) {
    y <- rnorm(n=2,mean=x[c("x1","x2")],sd=params["tau"])
    names(y) <- c("y1","y2")
    y
}

theta <- c(
    alpha.1=0.8, alpha.2=-0.5, alpha.3=0.3, alpha.4=0.9,
    sigma.1=3, sigma.2=-0.5, sigma.3=2,
    tau=1,
    x1.0=-3, x2.0=4
)

ou2 <- pomp(
    data=data.frame(
        time=1:100,
        y1=NA,
        y2=NA
    ),
    times="time",
    rprocess=discrete.time.sim(
        step.fun=ou2.proc.sim,
        delta.t=1
    ),
    rmeasure=ou2.meas.sim,
    t0=0
)
\end{verbatim}

\begin{figure}
\begin{center}
\includegraphics[width=14cm, height=12cm]{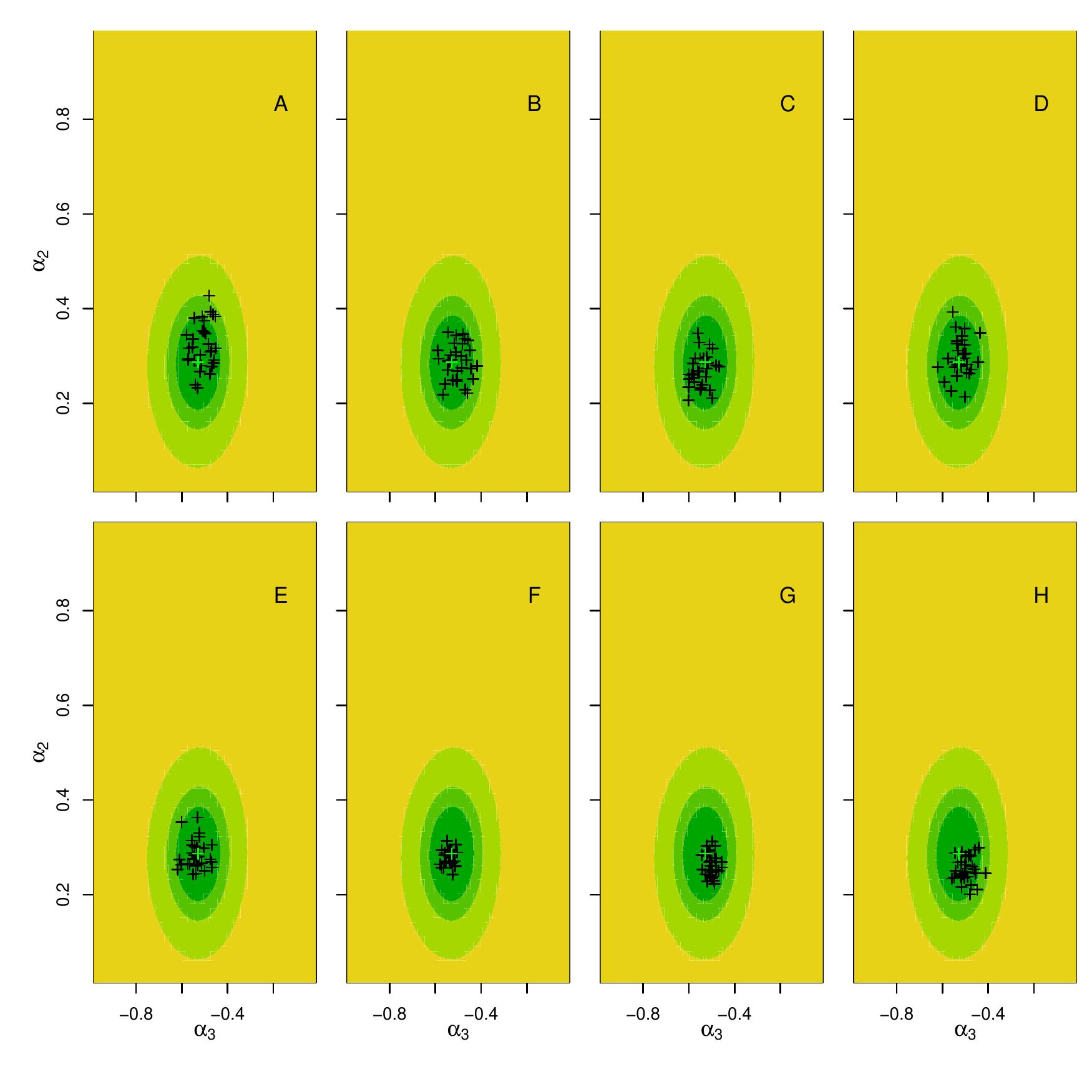}
\caption{
  Comparison of different estimators. The likelihood surface for the linear, Gaussian
model, with likelihood within $2$ log units of the maximum shown in green, within $4$ log units in light green, within $10$ log units in yellow green, and lower in yellow.
The location of the MLE is marked with a green cross. 
The black crosses show final points from 30 Monte Carlo replications of the estimators: (A) IF1 method; (B) IF2
method; (C) IS2 method; (D) AVIF method ; (E) AVIF method lag 1; (F) AIF method; (G) AIF lag 1 method; (H) Momentum IF method. 
Each method, was started uniformly over the rectangle shown, with $M=20$ iterations, $N=1000$ particles, and a random walk standard deviation decreasing from $0.02$ geometrically to $0.011$ for both
$\alpha_{2}$ and $\alpha_{3}$. 
}
\end{center}
\label{fig:linear}
\end{figure}

We simulate from these parameters to obtain our data for this model and plot to intuitively observe them. 
\begin{verbatim}
ou2 <- simulate(ou2,params=theta)

plot(ou2,variables=c("y1","y2"))
\end{verbatim}
We will estimate
the parameter based on our simulated data. For each method mentioned above, we estimate parameters $\alpha_{2}$
and $\alpha_{3}$ for this model using $J=1000$ particles and $M=20$ iterations. 
We start the initial search
uniformly on a rectangular region $[0,1]\times[-1,0]$. As
can be seen from Figure~\ref{fig:linear}, most of the replications
clustered near the true MLE (green cross) computed from Kalman filter,
while none of them stays in a lower likelihood region, implying they all successfully converge.
The results show that AIF is the most efficient method of all because
using AIF all the results clustered near the MLE and stayed in the highest likelihood region,
indicating a higher empirical convergence rate.
Algorithmically, AIF has similar computational costs with the other first
order approaches such as IF1, IF2, AVIF and is cheaper than the second order approach
IS2. In deed, additional overheads
for estimating the score function make the computation time of AIF a bit larger
compared to the computational time of IF1, IF2. However, with complex models
and large enough number of particles, these overheads become negligible
and the computational time of AIF will be similar to the other first order
approaches. The fact that they have the convergence rate of second order
with computation complexity of first-order shows that they are very
promising algorithms. In addition, the results show that they are all robust
to initial starting guesses.

\subsection{The Gompertz model}

\label{sec:gompertz:setup}

The Gompertz model postulates that the density, $X_{t+\dlta{t}}$,
of a population of organisms at time $t+\dlta{t}$ is determined by the
density, $X_{t}$, at time $t$ according to 
\begin{equation}
X_{t+\dlta{t}}=K^{1-e^{-r\,\dlta{t}}}\, X_{t}^{e^{-r\,\dlta{t}}}\,\varepsilon_{t}.\label{eq:gompertz1}
\end{equation}
In equation \eqref{eq:gompertz1}, $K$ is the handling of the population variable,
$r$ is a positive parameter, and the $\varepsilon_{t}$ are independent
and identically-distributed log-normal random variables with $\log\varepsilon_{t}\sim\normal(0,\sigma^{2})$.
In addition, suppose that the population density is measured
with some errors which are log-normally distributed: 
\begin{equation}
\log{Y_{t}}\;\sim\;\normal\left(\log{X_{t}},\tau^{2}\right).\label{eq:gompertz-obs}
\end{equation}
By transforming into logarithmic scale, we obtain
\begin{equation}
\log{X_{t+\dlta{t}}}\;\sim\;\normal\left(\left(1-e^{-r\,\dlta{t}}\right)\,\log{K}+e^{-r\,\dlta{t}}\,\log{X_{t}},\sigma^{2}\right).\label{eq:gompertz2}
\end{equation}

The step to setup gompertz is similar to ou2 model so we skip it and
use the default model from pomp by 
\begin{verbatim}pompExample(gompertz).
\end{verbatim} 
For Gompertz model, the parameters are passed in the argument \code{params} and
we will estimate them by \code{pmcmc} and \code{pmif} methods.
\begin{figure}
\centering
\includegraphics[height=9cm]{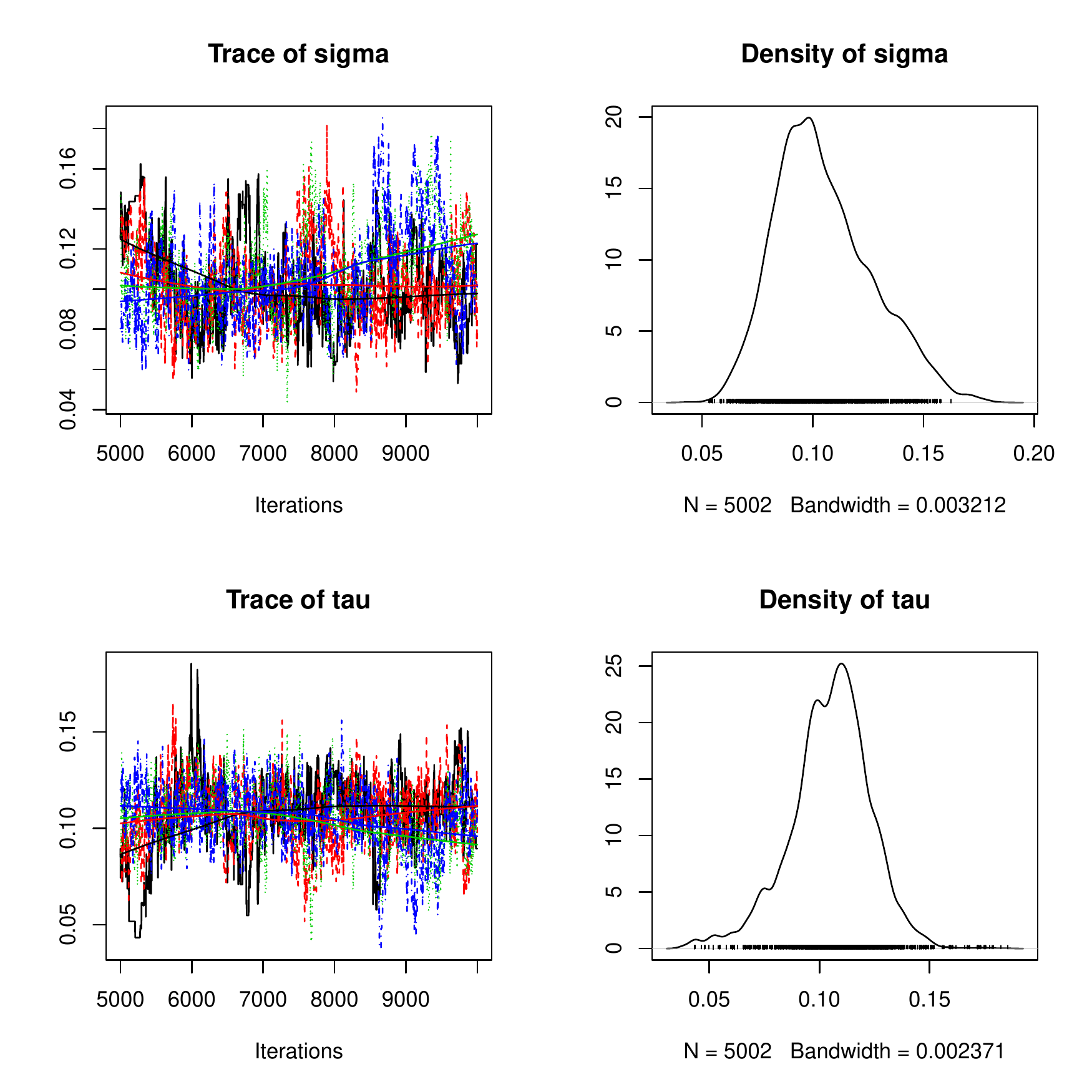}
\caption{
  Particle Markov chain Monte Carlo of the Gompertz model (\cref{eq:gompertz1,eq:gompertz-obs}).
  This figure shows the result of the 4 replicated chains of 10000 in length and the burn-in is 5000.
}
\label{fig:pmcmc-gompertz}
\end{figure}

\begin{figure}
\centering
\includegraphics[height=9cm]{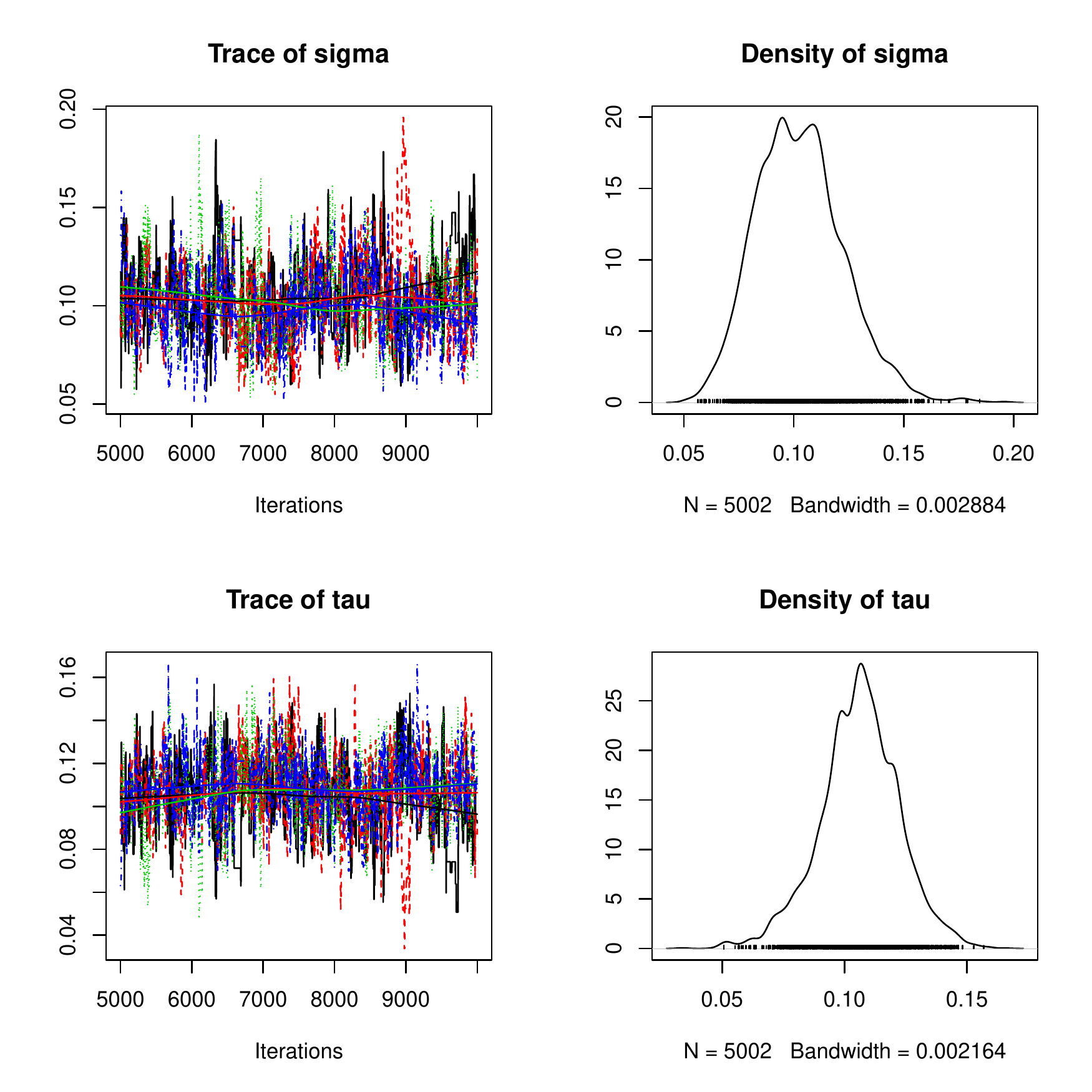}
\caption{
  Particle iterated filtering of the Gompertz model (\cref{eq:gompertz1,eq:gompertz-obs}).
  This figure shows the result of the 4 replicated chains of 10000 in length and the burn-in is 5000.
}
\label{fig:pmif-gompertz}
\end{figure}

%\caption{Computation times, in seconds, for the toy example.}
\begin{table}
\centering
\begin{tabular}{lrrrr}
\hline 
 & \rule[-1.5mm]{0mm}{7mm}\hspace{6mm}$ESS(\sigma)$ & \hspace{6mm}$ESS(\tau)$ & \tabularnewline
\hline 
PMCMC    & 240.4383 & 296.8016  & \tabularnewline
PMIF    & 395.3183 & 583.7596 & \tabularnewline
\hline 
\end{tabular}
\caption{Effective sample size for the gompertz example of 4 replications.}\label{table:2}
\end{table}

\begin{verbatim}
pmcmc <- foreach(
    i=1:4,
    .inorder=FALSE,
    .packages="is2",
    .combine=c,
    .options.multicore=list(set.seed=TRUE)
) %dopar% {
    pmcmc(pomp(gompertz, dprior = gompertz.dprior), start=coef(gompertz),
          Nmcmc = 10000, Np = 100, max.fail = Inf,
          proposal=mvn.diag.rw(c(r = 0.01, sigma = 0.01, tau = 0.01)))
}
######
pmif <- foreach(
    i=1:4,
    .inorder=FALSE,
    .packages="is2",
    .combine=c,
    .options.multicore=list(set.seed=TRUE)
) %dopar% {
    pmif(pomp(gompertz, dprior = gompertz.dprior), start=coef(gompertz),
          Nmcmc = 10000, Np = 100, max.fail = Inf,
          proposal=mvn.diag.rw(c(r = 0.01, sigma = 0.01, tau = 0.01)))
}
\end{verbatim}

It is well known that PMCMC does not mix well for some models (e.g. as observed in \citep{dahlin2015particle}). 
By providing extra information through iterated filtering, mixing of the Markov chain can improve significantly. 
To verify the claim, we compare the performance of PIF with original PMCMC by using
diagnostic plots and effective sample size function. Figures~\ref{fig:pmcmc-gompertz} and figures~\ref{fig:pmif-gompertz} show the resulting 5 Markov chains of 
the PMCMC and PIF algorithms respectively on the left panels and the posterior distributions of each estimated parameters on the right panels.  
While the diagnostic plot just gives us intuitively visual assessment of the mixing, the MCMC efficiency defined by the effective sample size function 
clearly demonstrates the effectiveness of \code{pmif} over \code{pmcmc} (Table~\ref{table:2}). 

\section{A more complex example}
\label{sec:EpidemicModel}
Many real world dynamic systems are highly nonlinear, partially observed, 
and even barely identifiable. To demonstrate the competencies of some of the implemented methodologies for such situations, we apply them to fit 
a malaria model in North-West India developed by \citet{roy13}. 
The reason to choose this challenging model
is that it provides a rigorous performance benchmark for our verification. 

We follow the setup of \citet{roy13} closely.
The model we examine divides the investigate population
of size $P(t)$ into distinct classes: susceptible individuals, $S(t)$, exposure $E(t)$,
infected individuals, $I(t)$, dormant classes $H_1(t)$, $H_2(t)$, $H_3(t)$  and recovered individuals, $Q(t)$.  
The last $S$ in the model name indicates the possibility that a recovered person can return to the class of susceptible individuals
since infection with malaria can lead to incomplete and waning immunity. 
Data, represented by $y_{1:N}^*$, are malaria morbidity reported each month. 
The state process is
\[\big(S(t),E(t),I(t),Q(t),H_1(t),H_2(t),H_3(t),\kappa(t), \mu_{SE}(t) \big),\] 
where $P(t)$ is from the census data while the birth rate for each class makes sure $S(t)+E(t)+I(t)+Q(t)+\sum_{i}H_{i}(t)=P(t)$ always satisfied. 
We suppose that $\{X(t),t\ge t_{0}\}$ follows
a stochastic differential equation in which the human stage of the malaria pathogen life-cycle is modeled by
\begin{eqnarray*}
dS/dt&=&\delta P+\mathrm{d}P/dt+\mu_{IS}I+\mu_{QS}Q+a\mu_{IH}I+b\mu_{EI}E-\mu_{SE}(t)S-\delta S,\\
dE/dt&=&\mu_{SE}(t)S-\mu_{EI}E-\delta E,\\
dI/dt&=&(1-b)\mu_{EI}E+3\mu_{HI}H_n
-(\mu_{IH}+\mu_{IS}+\mu_{IQ})I-\delta I,\\
dH_1/dt&=&(1-a)\mu_{IH}I-n\mu_{HI}H_1-\delta H_1,\\
dH_i/dt&=&3\mu_{HI}H_{i-1}-3\mu_{HI}H_i-\delta H_i\hspace{5mm}\mbox{for $i \in \{2,3\}$},\\
dQ/dt&=&\mu_{IQ}I-\mu_{QS}Q-\delta Q,
\end{eqnarray*}
where $\delta$ denote mortality rate. Specifically, it represents the average number of deceased people in that class per time unit. 
In this model, infected population enters dormancy via $I-\mathrm{to}-H$
transition at rate $\mu_{IH}$, and the treated humans join non-relapsing
infected in moving to the $\mathrm{Q}$ class. 
The transition rates from stage $H_1$ to $H_2$, $H_2$ to $H_3$ and $H_3$ to $Q$ are specified to be $3\mu_{HI}$.
The malaria pathogen reproduction within the mosquito vector is specified by
\begin{eqnarray*}
\mathrm{d}\kappa/\mathrm{d}t&=&[\lambda(t)-\kappa(t)]/\tau_D,\\
\mathrm{d}\mu_{SE}/\mathrm{d}t&=&[\kappa(t)-\mu_{SE}(t)]/\tau_D.
\end{eqnarray*}
The latent force of infection $\lambda(t)$ contributes to the current force of infection, $\mu_{SE}(t)$ with mean latency time $\tau_D$, after passing through a delay stage, $\kappa(t)$.
The relationship between $\lambda(t)$, $\kappa(t)$ and $\mu_{SE}(t)$ can be specified through
\begin{equation}
\mu_{SE}(t)= \int_{-\infty}^{t}\gamma(t-s)\lambda(s)\mathrm{d}s,
\end{equation}
with $ \gamma(s)=\frac{(2/\tau_D)^{2}s^{2-1}}{(2-1)!}\exp(-2s/\tau_D)$, a gamma distribution with shape parameter $2$.
As described by \citet{roy13}, the latent force of infection is given by 
\begin{equation*}
{\lambda(t)=\left(\frac{I+qQ}{P}\right)\times\exp\left\{\sum_{i=1}^{N_{s}}b_{i}s_{i}(t)+b_{r}R(t)\right\}{ \times\left[\frac{\mathrm{d}\Gamma(t)}{\mathrm{d}t}\right]}}\label{eq:lambda},
\end{equation*}
where $R(t)$ denotes rainfall covariate, $q$ denotes a reduced infection risk from humans in the $\mathrm{Q}$
class and  $\{s_{i}(t),i=1,\dots,N_{s}\}$ is a periodic cubic B-spline
basis, with $N_{s}=6$. Let  initial time $t_{0}=t_{1}-1/12$ and the system is measured at discrete time $t_1<t_2<\dots<t_N$ and the number of new cases in the $n$th interval 
be $M_{n}= \rho\int_{t_{n-1}}^{t_{n}}[\mu_{EI}E(s)+3\mu_{HI}H_{3}(s)]ds$. Also we assume that $Y_n$ given $M_n$ follows a negative binomial distribution with mean $M_n$ and variance $M_n+M_n^2\sigma^2_{\mathrm{obs}}$.
We use an Euler-Maruyama scheme \citep{kloeden99} with a time step of $1/20$ month to approximate the solution to the above coupled system of stochastic differential equations.

\begin{figure}\label{fig:malaria}
\begin{center}
\vspace{-0.5cm}
 \includegraphics[width=10cm,height=10cm]{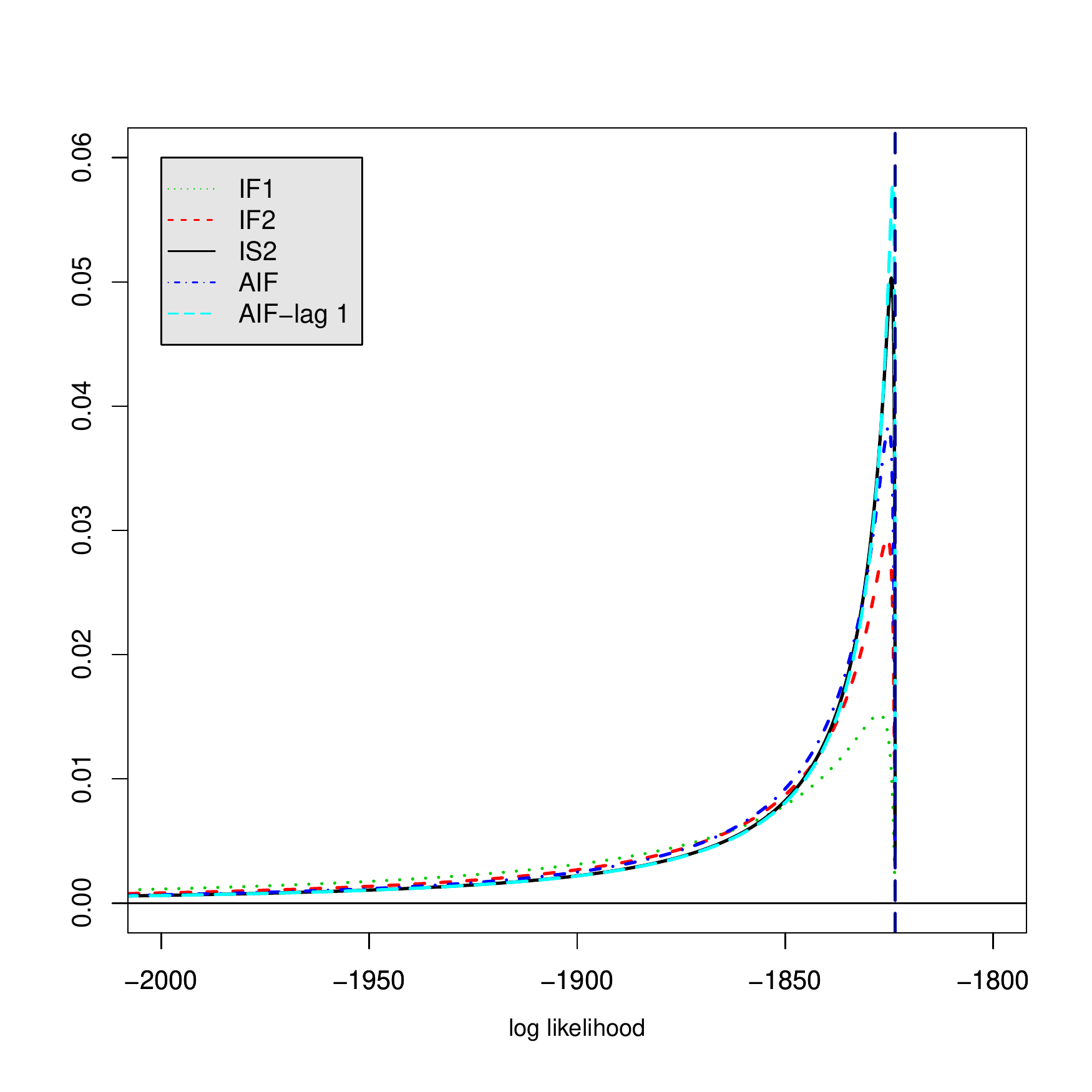} 
\caption{The density of the maximized log-likelihood approximations
estimated by IF1, IF2, IS2 and AIF for the malaria model when using
J = 1000 and M = 50. The log-likelihood at a previously computed
MLE is shown as a dashed vertical line.}
\end{center}
\label{fig:malaria}
\end{figure}
 \begin{table}
\begin{center}
\caption{Summary results of fitting malaria model using IF1, IF2, IS2 and AIF with the number of particle to $J=1000$ and 
the number of iteration $M=50$.}
\label{table:malaria}
\vspace{2mm}
\begin{tabular}{crrrrrr}
 \hline
 Algorithms& \rule[-1.5mm]{0mm}{7mm}\hspace{6mm} $\hat{\ell}$ &   \hspace{6mm} time(s)\\
  \hline
IF1  & -1904.418 &  19422.324\\ 
IF2  & -1879.970 &  22002.863\\
IS2  & -1870.557 &  23763.965\\
AIF  & -1861.709 &  22141.768\\
   \hline
\end{tabular}
\end{center}
\end{table}

We carried out simulation-based inference via the
original iterated filtering (IF1), the Bayes map iterated filtering
(IF2), second-order iterated smoothing (IS2) and the accelerated
iterated filtering (AIF) and AIF with lag 1. The inference goal used to assess all of these methods is to find
high likelihood parameter values starting from randomly drawn values
in a large hyper-rectangle. We provide this initial hyper-rectangle in the Supplement S-1. 
In the presence of possible
multi-modality, weak identifiability, and considerable Monte Carlo
error of this model, we start with about $50$ random searches. We use the standard setting as in \citep{nguyen2017second} to initially set the 
random walk standard deviation for estimated parameters to $0.1$ and the cooling rate $c$ to $0.1^{0.02}\approx0.95$. 
These corresponding quantities for initial value parameters are set to $2$ and
$0.1^{0.02}$, respectively. We run our experiment on a computer node with $M=50$ iterations
and with $J=1000$ particles. The chosen values are reasonable for exploring the likelihood surface of 
this model since increasing the iterations and the number
of particles does not improve the results much and it
significantly takes longer time. Figure~\ref{fig:malaria} shows the
the MLEs estimated by IF1, IF2, IS2 and AIF. Higher mean and smaller
variance of the MLEs, our implemented methods demonstrate that they are
considerably more effective than IF1. Table~\ref{table:malaria} shows the summary of the estimate likelihood and computational time.
Note that the computational times for IF1, IF2, IS2 and AIF are 19422.324, 22002.863, 23763.965 and 22141.768 seconds respectively, confirming
that all proposed methods have essentially the same computational
cost as the first order methods for a given Monte Carlo sample
size and number of iterations. While IF1 reveals
their limitations in this challenging problem, we have shown that IF2 and all new methods can still offer
a substantial improvement over IF1. Additional illustrations confirm it can be found online.

\section{Conclusion}\label{sec:conclusion}
\label{sec:conclusion}  The \pkg{is2} package is designed to be
a useful tool for time series modeling and analysis of POMP models
so that it is painless to make inference via open-source software.
Inherited robust platform from \pkg{pomp} package, a large set
of of inference are made available to practical researchers, freeing
them from writing their own code from scratch. Moreover, model specification
language focused on modeling domain, separating from the inference
method, facilitates model selection at the design stage. The examples
demonstrated in this paper are well represented the potential of the
package in a number of scientific studies.

As an open-source project, \pkg{is2} is especially convenient for
carrying out algorithms with the plug-and-play property, since models
will typically be specified by their \code{rprocess} simulator,
together with \code{rmeasure}. The package is easily extendable
for simulation and computationally efficient data analysis in various
ways: for example, different sampling could be added to the package.
In addition, while we focused on plug-and-play methodss, inference
approaches for non-plug-and-play can be added by building on generic
inference packages such as NIMBLE \citep{de2017programming} or STAN \citep{carpenter2017stan}. Advances in improving the
current class of inferences play an important part of the statistical
machinery for time series models, potentially applicable beyond
dynamic modeling.

Efficiency and scalability are necessary component of any modern software
packages. Currently, \pkg{is2} provides users with key functions written
in \proglang{C} and embarrassingly parallel computations. It is
possible to embed parallelization to increase the performance for
certain computationally intensive tasks but we leave it for our future
work.

To sum up, \pkg{is2} is currently effective for estimating numerous types
of infectious disease models, and the fixed lag smoothing component allows researchers
to employ the different sampling or resampling schemes in any desired models. As additional features are
augmented, we expect that the package keeps pace with \pkg{pomp} as a supplementary
set of tools for pomp computation and study. Last but not least, more
examples, which can be used as templates for implementation of new
models; the \proglang{R} and \proglang{C} code underlying these
examples are provided with the package. In addition, documentation
and an introductory vignette are provided with the package and on
the \pkg{is2} website \url{http://github.com/nxdao2000/is2}.

\section*{Acknowledgements}
This research is funded in part by the University of Mississippi Summer Grant.

\end{document}